\documentclass[a4paper,12pt]{article}
\usepackage{amssymb,amsmath}
\usepackage{graphics}
\def\beq{\begin{equation}}
\def\eeq{\end{equation}}
\def\bea{\begin{eqnarray}}
\def\eea{\end{eqnarray}}
\def\nn{\nonumber}

\makeatletter
  \def\@cite#1#2{${\mbox{#1\if@tempswa , #2\fi}}$}
\makeatother

\makeatletter
  \def\@biblabel#1{$^{\mbox{#1}}$}
\makeatother
\begin{document}
%%%%%%%%%%%%%%%%%%%%%%%%%%%%%%%%%%%%%%%%%%%%%%%%%%%%%%%%%%%%%%
%
%  title page
%
%
%%%%%%%%%%%%%%%%%%%%%%%%%%%%%%%%%%%%%%%%%%%%%%%%%%%%%%%%%%%%%%
%
\thispagestyle{empty}
\vspace*{3cm}
\begin{center}
{\LARGE\sf Extended Curie-Weiss law: a nonextensive perspective}\\

\bigskip\bigskip
R. Chakrabarti and R. Chandrashekar \\

\bigskip
\textit{
Department of Theoretical Physics, \\
University of Madras, \\
Guindy Campus, Chennai 600 025, India
}  

\end{center}

\vfill
\begin{abstract}
In the framework of the Tsallis nonextensive statistical mechanics we study an 
assembly of $N$ spins, first in a background magnetic field, and then assuming 
them to interact via a long-range homogeneous mean field. To take into account 
the spin fluctuations the dynamical field coefficient is considered to be 
linearly dependent on the temperature. The physical quantities are evaluated 
using a perturbative expansion in the nonextensivity parameter $(1-q)$. The 
extended Curie-Weiss law in the mean field case has been generalized. The 
critical temperature and the Curie-Weiss constant are found to be dependent on
the nonextensivity parameter $(1-q)$. 
\end{abstract}
PACS Number(s): 05.20.-y, 05.70  
\newpage
\setcounter{page}{1}
%%%%%%%%%%%%%%%%%%%%%%%%%%%%%%%%%%%%%%%%%%%%%%%%%%%%%%%%%%%%%%
%
%          Introduction 
%
%%%%%%%%%%%%%%%%%%%%%%%%%%%%%%%%%%%%%%%%%%%%%%%%%%%%%%%%%%%%%
%
\setcounter{equation}{0}
\section{Introduction}
\label{Intro}
A generalization of the Boltzmann-Gibbs extensive statistical mechanics 
was proposed by Tsallis [\cite{CT1}] via a deformation of the 
functional form of entropy
\beq
S = k \ln_{q} W,   \qquad   q \in {\mathbb{R}_{+}},
\label{Tqe}
\eeq
where $k$ is the Boltzmann constant, 
and $W$ is the weight factor. The stability of Tsallis entropy (\ref{Tqe})
is ensured by maintaining the deformation parameter $q$ as a positive real 
number [\cite{SA}]. 
The deformed $q$-logarithm and its inverse the $q$-exponential function read
\beq
\ln_{q} x = \frac{x^{1-q} -1}{1-q},  \qquad  
\exp_{q}(x) = [1+(1-q) x]^{\frac{1}{1-q}}.
\label{q_ln_exp}
\eeq
The  extensive classical Boltzmann-Gibbs entropy is recovered from (\ref{Tqe})
in the $q \rightarrow 1$ limit. Nonextensive statistical mechanics has found 
applications to a wide variety of 
fields such as anomalous diffusion [\cite{BMT},\cite{CRL}], quantum 
information theory [\cite{AR},\cite{BCPP}], astrophysical problems 
[\cite{OS},\cite{PHC}], and biological systems [\cite{HMO},\cite{MADG}].

\par

The introduction of physically appropriate constraints while the nonextensive 
entropy (\ref{Tqe}) of the system is maximized has been achieved in two 
alternate ways. In the so-called second constraint picture [\cite{CUR}] one 
works with the \textit {unnormalized} $q$-expectation values of a physical 
variable $O$:
\beq
\langle O\rangle_{q}^{(2)} = \sum_{j}\Big(
                             {\mathfrak{p}}_{j}^{(2)}(\beta)\Big)^{q}\,O_{j},
\qquad
\langle 1\rangle_{q}^{(2)} \equiv {\mathfrak{c}}^{(2)}(\beta)
                           = \sum_{j}\Big(
                             {\mathfrak{p}}_{j}^{(2)}(\beta)\Big)^{q},
\label{q_O_2}
\eeq
where $\mathfrak{p}_{j}^{(2)}(\beta)$ is the ensemble probability of the 
microstate $j$ in the second constraint picture: 
\beq
{\mathfrak{p}}_{i}^{(2)}(\beta) = \frac{\exp_{q}(- \beta E_{i})}
                                  {Z_{q}^{(2)} (\beta)},
\qquad
Z_{q}^{(2)}(\beta) = \sum_{i} \exp_{q}(- \beta E_{i}).
\label{ep_2c_def}
\eeq 
It is well-known that in the second constraint formalism the expectation 
value of the unit operator is \textit{not} preserved, and, as given 
in (\ref{q_O_2}), it equals to the sum of the $q$-weights 
$\mathfrak{c}^{(2)}(\beta)$.  
\par

In contrast the third constraint scenario [\cite{TMP}]     
employs appropriately \textit{normalized} escort probabilities resulting in 
the following $q$-expectation values:
\beq
\langle O\rangle_{q}^{(3)} = \frac{\sum_{j}
                             \Big({\mathfrak{p}}_{j}^{(3)}(\beta)\Big)^{q}
                             \,O_{j}}{{\mathfrak{c}}^{(3)}(\beta)},
\qquad
{\mathfrak{c}}^{(3)}(\beta) = \sum_{j} \Big( {\mathfrak{p}}_{j}^{(3)}
                              (\beta) \Big)^{q},
\label{q_O_3}
\eeq  
where ${\mathfrak{p}}_{j}^{(3)}(\beta)$ is the corresponding probability of 
the microstate $j$: 
\beq
{\mathfrak{p}}_{i}^{(3)}(\beta) = \frac{1}{{\bar{Z}_{q}^{(3)}}}
                                  \exp_{q}
                                  \left(- \beta \; \frac{E_{i} - U_{q}^{(3)}}
                                  {{\mathfrak{c}}^{(3)}(\beta)} \right),
\qquad
{\bar{Z}}_{q}^{(3)}(\beta) = \sum_{i} \exp_{q} 
                             \left(- \beta \; \frac{E_{i} - U_{q}^{(3)}}
                             {{\mathfrak{c}}^{(3)}(\beta)}\right),
\label{ep_3c_def}
\eeq
and ${\mathfrak{c}}^{(3)}(\beta)$ is the sum of $q$-weights. 
The thermodynamic averages of the physical quantities are obtained via the 
generalized partition function ${\bar{Z}}_{q}^{(3)}(\beta)$ that relates 
[\cite{TMP}] to the sum of the $q$-weights as 
\beq
 {\bar{Z}}_{q}^{(3)}(\beta) = \left({\mathfrak{c}}^{(3)}(\beta)\right)
                             ^{\frac{1}{1-q}}.
\label{ep_gpf}
\eeq
In both the above cases the inverse temperature $\beta \equiv (k T)^{-1}$ is 
associated [\cite{TMP}] with the Lagrange multiplier corresponding to the 
internal energy. In the light of the fact that in the third constraint 
formalism the unit operator trivially preserves its norm for an arbitrary $q$, 
it is considered [\cite{TMP}] to be fully satisfactory. Fortunately, the 
ensemble probabilities in the two pictures may be interrelated [\cite{TMP}] by 
the following equivalence relation:
\beq
{\mathfrak{p}}_{i}^{(3)}(\beta) = {\mathfrak{p}}_{i}^{(2)}(\beta^{\prime})
\label{ep_eq_rel}, 
\eeq
where the general recipe for constructing the transformation [\cite{TMP}] 
to the auxiliary temperature $\beta^{\prime}$ reads
\beq
\beta = \beta^{\prime} \frac{{\mathfrak{c}}^{(2)}(\beta^{\prime})}
        {1- (1-q) \beta^{\prime} \frac{U_{q}^{(2)}(\beta^{\prime})}
        {{\mathfrak{c}}^{(2)}(\beta^{\prime})}},
\qquad
\label{trans}
\eeq
where $U_{q}^{(2)}(\beta)$ is the internal energy in the second constraint 
picture. Its explicit computation pertinent to our models will be discussed 
later. Here we note that as a result of the above equivalence property 
the dynamical quantities obtained in the second constraint framework may be 
translated to their respective values corresponding to the choice of the 
third constraint.

\par

Application of nonextensive statistical mechanics to spin systems  
was initiated in [\cite{PPT}], where an assembly of $N$ noninteracting 
spin-$\frac{1}{2}$ particles in a background field was studied in the 
second constraint formalism, associating the inverse temperature with the 
Lagrange multiplier corresponding to the energy. A similar study based on 
the third constraint framework was also performed [\cite{MPP}]. The magnetic 
susceptibility in this model showed [\cite{PPT},\cite{MPP}] the interesting 
feature referred to as {\it dark magnetism}, indicating that the apparent 
number of spins are different from the actual number of spins. Employing a 
high temperature limit, it was observed in [\cite{MPP}] that in the domain 
$q > 1\; (q < 1)$ the effective number of spins $N_{\rm{eff}} > N \;
(N_{\rm{eff}} < N)$. In the study of manganites nonextensive statistical 
mechanics has been observed [\cite{RFOLO}] to fit the experimental data on 
magnetization better than the standard Boltzmann-Gibbs statistics.  A 
numerical analysis of a multilevel spin model has been done [\cite{RAALO}] 
following Tsallis statistics. A collection of spin clusters has been examined 
[\cite{RAAO1},\cite{RASO}] in the third constraint picture.
These authors obtained generalized paramagnetic susceptibility in the 
noninteracting regime, and a nonextensive modification of the Curie-Weiss law 
in the context of the mean field model.

\par

In our current work we make a slight departure from the Refs. 
[\cite{PPT},\cite{MPP}]. We examine a classical arbitrary $N$-spin system in 
a weak background magnetic field \textit{without} adopting the high 
temperature limit. The thermodynamic quantities in the second and the 
third constraint pictures are evaluated as a perturbative series in the 
nonextensivity parameter $(1-q)$ by disentangling the $q$-exponential 
(\ref{q_ln_exp}). This process of perturbative expansion, that may be continued
to an arbitrary order, is based on the technique developed in [\cite{CQ}] and 
previously used in [\cite{CCN1},\cite{CCN2}]. In passing from the second 
constraint picture to that of the third constraint we employ a 
transformation [\cite{CCN2}] procedure that allows us to express the physical 
quantities in the later scenario directly in terms of the former. In our 
perturbative expansion for the spins in the background field we retain terms 
at all orders of temperature. Subsequently we study interacting spins in an 
extended form of the mean field model [\cite{ZXXZ}] where the 
field strength coefficient (the proportionality factor) is made temperature 
dependent to accommodate quantum fluctuations among allowed configurations 
[\cite{SW}].  The critical temperature and the Curie-Weiss constant have been 
evaluated in the third constraint framework by retaining terms in the 
perturbation scheme up to the order $(1 - q)^2$. In particular, the critical
temperature in the nonextensive regime increases (decreases) compared to its 
value given by the Boltzmann-Gibbs statistics for the domain $q > 1\; (q < 1)$.
Our observation qualitatively agrees with the results obtained in 
[\cite{RAAO}] where a different definition of temperature for the nonextensive 
spin system has been used.  
The plan of this article is 
as follows: Spins in the 
presence of a weak external magnetic field is considered in Sec. \ref{WFL}. 
This is followed by the consideration of an extended mean field model in 
Sec. \ref{MFM}. Our concluding remarks are given in Sec. \ref{REM}.

%%%%%%%%%%%%%%%%%%%%%%%%%%%%%%%%%%%%%%%%%%%%%%%%%%%%%%%%%%%%%%%%%%%%%%%%
%
% Noninteracting spins in the weak field limit
%
%%%%%%%%%%%%%%%%%%%%%%%%%%%%%%%%%%%%%%%%%%%%%%%%%%%%%%%%%%%%%%%%%%%%%%%%
%
\setcounter{equation}{0}
\section{Spins in a  weak background field}
\label{WFL}
The classical Hamiltonian of a system of $N$ spins in the 
presence of a background magnetic field $H$ is given by 
\beq
E =  - \mu H  \sum_{i = 1}^{N} \cos \theta_{i},
\label{eev}
\eeq
where $\mu$ is the magnetic moment of the spins oriented at polar angles 
$(\theta_{i}, \phi_{i}| i = 1,\ldots, N)$ with the field. The partition 
function of the system in the second constraint picture reads
\beq
Z_{q}^{(2)}\left( \beta\right) = \int_{\theta_{i}=0}^{\pi}
                                 \int_{\phi_{i}=0}^{2\pi}
				 \left[ 1+(1-q)\beta \mu H \sum_{i=1}^{N}  
                                 \cos \theta_{i} \right]^{\frac{1}{1-q}}
                                 \prod_{i=1}^{N}
				 \sin \theta_{i}\, \hbox{d} \theta_{i}  \
				 \hbox{d} \phi_{i}.
				 \label{pfi_2c}
\eeq
In contrast to the extensive case the available phase space of integration in 
(\ref{pfi_2c}) depends on the strength of the magnetic field.    
In the regime of weak magnetic field  
\beq
\hat{\beta} \equiv \beta \mu H < \frac{1}{|1-q|\;N}, 
\label{H_lim}
\eeq
the integrand is real and positive in the whole phase space, and the partition 
function may be obtained as follows: 
\beq
Z_{q}^{(2)}\left( \beta\right) = (2\pi)^{N} \;\;\Phi\;\;
                                 \hat{\beta}^{-N}
				 {\mathfrak{S}}_{1}(\hat{\beta}), \qquad
\Phi = \prod_{\ell=1}^{N} \left[ 1 + (1-q) \ell \right]^{-1},
\label{pf_2c_wf}
\eeq
where the binomial sum ${\mathfrak{S}}_{1}(x)$ involving the $q$-exponentials 
is given by
\beq
{\mathfrak{S}}_{1}(x)  = \sum_{n=0}^{N} \binom{N}{n} (-1)^{n} \;\;
                            \left( \exp_{q} ((N-2n) x) \right)
		            ^{\Lambda},\qquad \Lambda = 1+(1-q)N.
\label{def_s1}
\eeq
In the extensive $q\rightarrow 1$ limit above sum 
approaches its well-known classical value
${\mathfrak{S}}_{1}(x) \rightarrow 2^{N} \sinh^{N} x$.
The expression (\ref{pf_2c_wf}) for the partition function  exhibits 
singularities at $q = 1 + \frac{1}{n}$ for $n = 1,\ldots,N$. 
These singularities are similar to the ones observed in 
[\cite{CCN1},\cite{CCN2}], and reflect the fact that the number of degrees of 
freedom plays a key role in determining the allowed range of the 
nonextensivity parameter $q$. 

\par

The internal energy in the second constraint picture [\cite{CUR}] 
\beq
U_{q}^{(2)}(\beta) = - \frac{\partial}{\partial\,\beta}\;
                       \ln_{q}\,Z_{q}^{(2)}(\beta)
\label{Ie_2c}
\eeq
is evaluated by employing the partition function (\ref{pf_2c_wf}):
\beq
U_{q}^{(2)}(\beta)  = \mu H \;  {\mathfrak{N}} \; \;
                      ({\mathfrak{S}}_{1}(\hat{\beta}))^{-q}\;
                      \hat{\beta}^{(q-1)N}
		      \left( N  \hat{\beta}^{-1}\; 
		      {\mathfrak{S}}_{1}(\hat{\beta})
		      - {\mathfrak{S}}^{\prime}_{1}(\hat{\beta})\right),
\label{Ie_2c_wf}
\eeq
where
\beq
{\mathfrak{N}} =  (2 \pi)^{(1-q)N} \Phi^{1-q}.
\label{N_scale}
\eeq
Here and elsewhere the primed functions indicate derivatives with respect to 
their arguments. The corresponding magnetization is obtained via 
the defining relation 
[\cite{PPT}] 
\beq
M_{q}^{(2)} \equiv \frac{1}{\beta} \frac{\partial}{\partial H} 
              \ln_{q} Z_{q}^{(2)}.
\label{mag_def}
\eeq
On subsequent use of a $q$-deformed Langevin function 
$L_{q}(x)$ the magnetization (\ref{mag_def}) assumes the form
\beq
M_{q}^{(2)} = \mu \; N \; {\mathfrak{N}}\; L_{q}(\hat{\beta}),
\label{mg_2c_wf}
\eeq
where
\beq
L_{q}(x) =  x^{(q-1)N} \; ({\mathfrak{S}}_{1}(x))^{1-q} \;
           \left( \coth_{q}(x;N) - x^{-1} \right), \qquad
\coth_{q}(x;N) =  \frac{{\mathfrak{S}}^{\prime}_{1}(x)}
                  {N \; {\mathfrak{S}}_{1}(x)}.
\label{def_lf}
\eeq
In the extensive $q \rightarrow 1$ limit, the magnetization (\ref{mg_2c_wf})
reduces to its well-known classical value.  The magnetization in the second
constraint as a perturbative series in $(1-q)$ reads
\beq
M_{q}^{(2)} = \mu \; {\mathfrak{N}} \;
              \Big(L(\hat{\beta}) + (1-q) \; {\mathsf{M}}_{1}+(1-q)^{2} \;
              {\mathsf{M}}_{2} + \ldots\Big),
\quad
L(x) = \coth(x)-x^{-1}.
\label{mg_2c_ps}
\eeq
The coefficients in the perturbative series up to $(1 - q)^{2}$ are given by
\bea
{\mathsf{M}}_{1} &=&  N^{2} \; L(\hat{\beta}) \; 
                     \ln {\mathcal{Z}}({\hat{\beta}})
                     -N(1-N) \; {\hat{\beta}}^{2} \coth {\hat{\beta}} \;
                     {\hbox{cosech}}^{2} {\hat{\beta}}+  N^{2}
                     \coth^{2} {\hat{\beta}}    \nn \\
                 & &   + N(1-2N) {\hat{\beta}} \coth^{2} {\hat{\beta}} 
                       - N(1-N) {\hat{\beta}}, \nn \\
{\mathsf{M}}_{2} &=& -N^{2}(1-2N) {\hat{\beta}} + N(1-N) {\hat{\beta}}^{3}
                      - 6N(1-N+N^{2}) 
                     {\hat{\beta}}^{3} \coth^{2} {\hat{\beta}} \;
                     {\hbox{cosech}}^{2} {\hat{\beta}}  \nn \\ 
                 & & -N^{3} \coth {\hat{\beta}} - (N-5 N^{2} \; {\hat{\beta}})
                     {\hat{\beta}}^{3} \coth^{4} {\hat{\beta}} +\frac{N}{2}
                     (1- N-12N \; {\hat{\beta}}^{2}) 
                     {\hat{\beta}} \coth^{2} {\hat{\beta}} \nn \\
                 & & + \frac{N^{3}}{2} {\hat{\beta}}^{2} \coth^{3} 
                     {\hat{\beta}} + N (3-5N+2N^{2}) {\hat{\beta}}^{4}
                     \coth^{3} {\hat{\beta}} \; 
                     {\hbox{cosech}}^{2} {\hat{\beta}}
                     \nn \\ 
                & &  +\frac{N}{2} \left((4-13N+10N^{2}) -
                     (4-6N+2N^{2}) {\hat{\beta}}^{2} \right) {\hat{\beta}}^{2}
                     \coth {\hat{\beta}} \;
                     {\hbox{cosech}}^{2}{\hat{\beta}} \nn 
                     \\
                & &  +\frac{N}{3} (2-6N+3N^{2}) {\hat{\beta}}^{3} 
                     {\hbox{cosech}}^{2} {\hat{\beta}} + \frac{N^{3}}{2}
                     (\ln {\mathcal{Z}}(\hat{\beta}))^{2} L(\hat{\beta}) \nn \\
                & &  -N^{3} {\hat{\beta}} \; \coth {\hat{\beta}} \;
                     \ln {\mathcal{Z}} (\hat{\beta}) \; L(\hat{\beta}) 
                     +N^{2}(1-N) {\hat{\beta}} \; {\hbox{cosech}}^{2} 
                     {\hat{\beta}} \; \ln {\mathcal{Z}} (\hat{\beta})  \nn \\
                & &  +N^{2} {\hat{\beta}}^{2} \coth {\hat{\beta}} \;
                     {\hbox{cosech}}^{2} {\hat{\beta}} \; \ln {\mathcal{Z}}
                     (\hat{\beta}), 
\label{mg_2c_ct}
\eea
where ${\mathcal{Z}}(x) = 2 \sinh(x)/x$. The perturbative evaluation of the 
internal energy in the second constraint framework is found to follow the 
standard thermodynamic relation
\beq
U = - M \; H.
\label{ie_mg_rel}
\eeq
For the sake of brevity we refrain from quoting it explicitly.

\par

Our task now is to translate the previous results to the third constraint 
picture. As evident from the context of (\ref{trans}) the sum of the 
$q$-weights plays a seminal role in enacting this transformation. The 
definitions (\ref{q_O_2}) and (\ref{ep_2c_def}) lead to the relation 
\beq
{\mathfrak{c}}^{(2)}(\beta) = \varOmega(\beta)\;\; (Z_{q}^{(2)})^{-q}.
\label{qw_2c_wf}
\eeq
In the above equation the sum $\varOmega(\beta)$ is given by 
\beq
\varOmega(\beta) \equiv \sum_{i} \left[ 1-(1-q) \beta E_{i} \right]
  ^{\frac{q}{1-q}}
= (2 \pi)^{N} \;\; \Phi \;\; \Lambda \;\; 
             \hat{\beta}^{-N} \;\; {\mathfrak{S}}_{2}(\hat{\beta}),  
\label{qw_nr}             
\eeq
where the binomial sum ${\mathfrak{S}}_{2}(x)$ reads 
\beq
{\mathfrak{S}}_{2}(x) = \sum_{n=0}^{N} \binom{N}{n} (-1)^{n} \;\;
                       \left(\exp_{q}((N-2n) x) \right)^{\Lambda-(1-q)}.
\label{def_s3}
\eeq
Employing (\ref{qw_nr}) and (\ref{pf_2c_wf}) we may express the sum of 
$q$-weights in the second constraint picture as 
\beq
{\mathfrak{c}}^{(2)}(\beta) = {\mathfrak{N}} \;\; 
                             \Lambda \;\;
                             \hat{\beta}^{(q-1)N} \;\; 
                             {\mathfrak{S}}_{2}(\hat{\beta}) \;\;
                             ({\mathfrak{S}}_{1}(\hat{\beta}))^{-q},
\label{qwfe_2c_wf}
\eeq
and its perturbative evaluation up to terms $O(1-q)^{3}$ reads:  
\beq
{\mathfrak{c}}^{(2)}(\beta) = {\mathfrak{N}} \;\;  
                              \left(1+(1-q) \; {\mathcal{P}}_{1} +
                              (1-q)^{2} \; {\mathcal{P}}_{2} + 
                              (1-q)^{3} \; {\mathcal{P}}_{3}+ \ldots\right),
\label{qw_2c_ps}
\eeq
where the perturbative coefficients may be enlisted as
\bea
{\mathcal{P}}_{1} &=& N \ln {\mathcal{Z}}(\hat{\beta}) - N {\hat{\beta}}
                      \coth {\hat{\beta}} + N, \nn \\
{\mathcal{P}}_{2} &=& \frac{N^{2}}{2} {\hat{\beta}}^{2} \; 
                      \coth^{2} {\hat{\beta}}-\frac{N}{2} 
                      \left((1-2N) - 2(1-N) {\hat{\beta}} \;
                      \coth {\hat{\beta}}\right) {\hat{\beta}}^{2} \;
                      {\hbox{cosech}}^{2} {\hat{\beta}} \nn \\
                  & & +\frac{N^{2}}{2} \left(2 -2 {\hat{\beta}} \; 
                      \coth {\hat{\beta}}
                      +\ln {\mathcal{Z}} (\hat{\beta})
                      \right) \;
                      \ln {\mathcal{Z}}(\hat{\beta}), \nn \\
{\mathcal{P}}_{3} &=& N^{3} {\hat{\beta}} \; (1+4 \,{\hat{\beta}}^{2})\; 
                      \coth {\hat{\beta}} + N (3-5N-2N^{2}) \;
                      {\hat{\beta}}^{5} \coth^{3} {\hat{\beta}} 
                      {\hbox{cosech}}^{2} {\hat{\beta}} \nn \\
                  & & +\frac{N}{12}(17-33N+12N^{2}) {\hat{\beta}}^{4}
                      {\hbox{cosech}}^{2} {\hat{\beta}} 
                      - \frac{N^{3}}{2} \;{\hat{\beta}}^{2}
                      (1+\coth^{2} {\hat{\beta}}) \nn \\
                  & & + \frac{N}{4} (17-39N+22 N^{2}) {\hat{\beta}}^{4} \,
                      \coth^{2} {\hat{\beta}} \; {\hbox{cosech}}^{2}
                      {\hat{\beta}}  \nn \\
                  & & +\frac{N^{3}}{6} \ln {\mathcal{Z}} (\hat{\beta})
                      \left( 3 \, {\hat{\beta}}^{2} 
                      \coth^{2}{\hat{\beta}} 
                      +3(1-{\hat{\beta}} \coth {\hat{\beta}})
                      \ln {\mathcal{Z}}(\hat{\beta}) + 
                      (\ln {\mathcal{Z}} (\hat{\beta}))^{2}  \right) \nn \\
                  & & - \frac{N^{2}(1-N)}{2} (1-2 {\hat{\beta}}
                      \coth {\hat{\beta}}) {\hat{\beta}}^{2} \;
                      {\hbox{cosech}}^{2} {\hat{\beta}} \;
                      \ln {\mathcal{Z}} (\hat{\beta}). 
\label{qw_2c_pt}
\eea
Following an approach used in [\cite{CCN2}] we interrelate the dynamical 
quantities such as the internal energy and the magnetization directly from the 
second to the third constraint picture. The equivalence of the ensemble 
probabilities (\ref{ep_eq_rel}) leads to a ready translation of 
the expectation values of an observable $O$ in the two constraint frameworks 
as  
\beq
O_{q}^{(3)} (\beta) = \frac{O_{q}^{(2)}(\beta^{\prime})}
                      {{\mathfrak{c}}^{(2)} (\beta^{\prime}) }.
\label{Otrans_23}
\eeq
To fruitfully employ this procedure we need to invert the transformation 
relation (\ref{trans}) of the temperature. As a closed form inversion rule is 
not at hand, we adopt a perturbative technique [\cite{CCN1}], and express the
auxiliary temperature $\beta^{\prime}$ in terms of the physical temperature 
$\beta$ retaining terms up to second order in $(1-q)$: 
\beq
\beta^{\prime} = \frac{\beta} {{\mathfrak{N}}} 
                 \left(1+(1-q) \;  g(\overline{\beta}) + 
                 (1-q)^{2} \; h(\overline{\beta})+\ldots \right),
\label{bbp_tr} 
\eeq
where the rescaled temperature is given by ${\overline{\beta}} = \hat{\beta}\,
{{\mathfrak{N}}}^{-1}$. The perturbative terms of the transformation 
relation read
\bea
g(\overline{\beta}) &=& -2N \, (1-{\overline{\beta}} \; 
                        \coth {\overline{\beta}}) - N 
                        \ln {\mathcal{Z}}({\overline{\beta}}), \nn \\
h(\overline{\beta}) &=& -\frac{3N}{2}\, {\overline{\beta}}^{2} (1-\coth^{2}
                        {\overline{\beta}}) - 2N(1+N)\,{\overline{\beta}}^{3}
                       \; \coth {\overline{\beta}} \; {\hbox{cosech}}^{2}
                        {\overline{\beta}} - 5 N^{2}\, {\overline{\beta}} \;
                        \coth {\overline{\beta}} \nn \\
                    & & + \frac{5N^{2}}{2} \, {\overline{\beta}}^{2} 
                        \coth^{2} {\overline{\beta}} + 3 N^{2} \,
                        {\overline{\beta}}^{2} \; {\hbox{cosech}}^{2} 
                        {\overline{\beta}} - N^{2} \, (1-{\overline{\beta}}^{2}
                         ) + \frac{N^{2}}{2} \, (\ln {\mathcal{Z}} 
                        (\overline{\beta}))^{2} \nn \\
                    & & -3N^{2} \; {\overline{\beta}} \;
                        \coth {\overline{\beta}} \;
                        \ln {\mathcal{Z}} ({\overline{\beta}}) + 2 N^{2}
                        {\overline{\beta}}^{2} {\hbox{cosech}}^{2}
                        {\overline{\beta}} \;
                        \ln {\mathcal{Z}} (\overline{\beta}).
\label{tr_pt}
\eea
Aided by the inverse transformation series (\ref{bbp_tr}) we employ 
(\ref{Otrans_23}) for obtaining a perturbative evaluation of the magnetization 
in the third constraint approach to the order $O((1-q)^{2})$:
\beq
M_{q}^{(3)} (\beta) = \mu \; \left( N \; L({{\overline{\beta}}}) 
                      + (1-q) \; {\mathfrak{M}}_{1} + (1-q)^{2} \;
                      {\mathfrak{M}}_{2} + \ldots\right),
\label{mg_3c_ps}
\eeq
where the coefficients ${\mathfrak{M}}_{1}$ and ${\mathfrak{M}}_{2}$ read
\bea
{\mathfrak{M}}_{1} &=& N^{2}\; (2 {\overline{\beta}}^{2} -1) \;
                       L({\overline{\beta}})
                       - N^{2} \; {\overline{\beta}} \; \ln {\mathcal{Z}}
                       ({\overline{\beta}}) \;
                       L^{\prime} ({\overline{\beta}}), 
                       \nn \\
{\mathfrak{M}}_{2} &=& 2 N^{3} \; L({\overline{\beta}}) + 4 N^{3} \;
                       {\overline{\beta}} - \frac{N^{2}}{2} \;
                       {\overline{\beta}}^{3} - N \; {\overline{\beta}}^{3}
                       -(2N+5N^{2}) \; {\overline{\beta}}^{3} \; \coth^{2}
                       {\overline{\beta}} \; {\hbox{cosech}}^{2} 
                       {\overline{\beta}} \nn \\
                   & & + \frac{N}{2} \; \left(4+10N-(4+9N^{2}) \; 
                       {\overline{\beta}}^{2}\right)
                       {\overline{\beta}}^{2} \;
                       \coth {\overline{\beta}} \; {\hbox{cosech}}^{2}
                       {\overline{\beta}} \nn \\
                   & & + 2N(6+N+N^{2})\, {\overline{\beta}}^{4} \coth^{3} 
                       {\overline{\beta}} \; {\hbox{cosech}}^{2} 
                       {\overline{\beta}} - \frac{N}{2} (9N^{2} + 4N
                       {\overline{\beta}} - 8 {\overline{\beta}}^{2})
                       {\overline{\beta}} \coth^{2} {\overline{\beta}} 
                       \nn \\
                   & & - \frac{3N}{2} (2+N) \, {\overline{\beta}}^{3} \;
                       \coth^{4} {\overline{\beta}} +\frac{N^{2}}{6} \;
                       \left(9-36N - (4-6N) \;
                       {\overline{\beta}}^{2}\right) {\overline{\beta}}\;
                       {\hbox{cosech}}^{2} {\overline{\beta}} \nn \\
                   & & +N \; {\overline{\beta}}^{4} \;
                       \coth {\overline{\beta}} \;
                       {\hbox{cosech}}^{4} {\overline{\beta}}.
\label{mg_3c_ct}
\eea
The internal energy $U_{q}^{(3)}(\beta)$ in the third constraint picture may 
be read off directly from the corresponding magnetization $M_{q}^{(3)}(\beta)$ 
via the general thermodynamic relation (\ref{ie_mg_rel}). We do not quote it 
explicitly. Magnetic susceptibility is defined as
\beq
\chi_{q}^{(3)} (\beta)  \equiv \frac{\partial M_{q}^{(3)} }{\partial H}
\label{ms_def},
\eeq
and we now employ (\ref{mg_3c_ps}) to compute it: 
\beq
\chi_{q}^{(3)} (\beta) = \frac{\mu^{2}\,\beta}{\mathfrak{N}} 
                         \left(\left(\frac{N}{\overline{\beta}} \right)^{2}
                          -N \,{\hbox{cosech}}^{2} {\overline{\beta}} + 
                          (1-q) \;  \vartheta_{1} + (1-q)^{2} \;
                          \vartheta_{2} + \ldots\right).
\label{ms_3c_ps}
\eeq
The above perturbative coefficients up to the 
order $O((1-q)^2)$ are given below:  
\bea
\vartheta_{1} &=& \frac{N^{2}}{{\overline{\beta}}^{2}} \ln {\mathcal{Z}}
                  ({\overline{\beta}}) + N \left(1 + N + N \ln {\mathcal{Z}}
                  ({\overline{\beta}})\right) 
                  {\hbox{cosech}}^{2} {\overline{\beta}}+\frac{2 N^{2}}
                  {{\overline{\beta}}^{2}} - 2N^{2} {\hbox{cosech}}^{2}
                  {\overline{\beta}} \nn \\
              & & +N(1+N){\overline{\beta}}^{2} \, {\hbox{cosech}}^{4} 
                  {\overline{\beta}} + 2N (1+N) {\overline{\beta}}^{2} 
                  \coth^{2} {\overline{\beta}} \, {\hbox{cosech}}^{2} 
                  {\overline{\beta}} - \frac{N^{2}}{\overline{\beta}}
                  \coth {\overline{\beta}} \nn \\
              & & - N \, {\overline{\beta}} \left(4N^{2} + \ln {\mathcal{Z}}
                  ({\overline{\beta}})\right)
                  \coth {\overline{\beta}} \, {\hbox{cosech}}^{2} 
                  {\overline{\beta}}, \nn \\
\vartheta_{2} &=& \frac{4 N^{3}} {{\overline{\beta}}^{2}} +
                  \frac{3 N^{3}} {\overline{\beta}} \coth {\overline{\beta}}
                  + \frac{N^{2}}{2} \left(N \coth^{2} {\overline{\beta}}
                  - {\hbox{cosech}}^{2} {\overline{\beta}}\right) 
                  +N(4+N+17N^{2}){\overline{\beta}} \,
                  \coth {\overline{\beta}} \nn \\
              & & +\frac{N}{2} \left((6-5N+30N^{2}) \;
                  {\hbox{cosech}}^{2} {\overline{\beta}} - (18+47N^{2})
                  \coth^{2} {\overline{\beta}}\right)  
                  {\overline{\beta}}^{2} \; {\hbox{cosech}}^{2} 
                  {\overline{\beta}} \nn \\ 
              & & + 4 N^{2} \left(N-1+(1+2N) \coth^{2} {\overline{\beta}}
                  \right) {\overline{\beta}}^{3} \coth {\overline{\beta}}
                  {\hbox{cosech}}^{2} {\overline{\beta}} - N^{2}(1-N)
                  {\overline{\beta}}^{4} {\hbox{cosech}}^{2} {\overline{\beta}}
                  \nn \\
              & & + N^{3} (7+22N) \,{\overline{\beta}} \; 
                  \coth {\overline{\beta}}\;
                  {\hbox{cosech}}^{2} {\overline{\beta}} 
                  - N (2+3N+3N^{2}) \,
                  {\overline{\beta}}^{4} 
                  \coth^{4} {\overline{\beta}} \, {\hbox{cosech}}^{2} 
                  {\overline{\beta}} 
                \nn \\
              & & - N \left((12+25N-19N^{2})\coth {\overline{\beta}} + 
                  N(3-13N)\right)  
                  {\overline{\beta}}^{2} \coth {\overline{\beta}}\, 
                  {\hbox{cosech}}^{2} {\overline{\beta}} \nn \\ 
              & & - \frac{N}{3}
                  \left(40-39N-48N^{2}-(32-54N+15N^{2})
                  \coth^{2} {\overline{\beta}} \right) 
                  {\overline{\beta}}^{3} 
                  \coth {\overline{\beta}} \, {\hbox{cosech}}^{2} 
                  {\overline{\beta}} \nn \\
              & & +N \big(2N(1+N)  
                  + (13-12N-7N^{2})  \coth^{2}
                  {\overline{\beta}}\big) \, {\hbox{cosech}}^{4} 
                  {\overline{\beta}} \nn \\
              & & + \frac{3 N^{3}} 
                  {{\overline{\beta}}^{2}} \ln {\mathcal{Z}}
                  ({\overline{\beta}}) - \frac{N^{3}}{{\overline{\beta}}}
                  \coth {\overline{\beta}} \, \ln {\mathcal{Z}}
                  ({\overline{\beta}}) \nn \\
              & & -N^{2}(1-N) \, {\hbox{cosech}}^{2} {\overline{\beta}} \,
                  \ln {\mathcal{Z}}({\overline{\beta}})  -N^{2}(10-N) \, 
                  {\overline{\beta}} \, \coth {\overline{\beta}} \,
                  {\hbox{cosech}}^{2} {\overline{\beta}} \, \ln {\mathcal{Z}}
                  ({\overline{\beta}}) \nn \\
             & &  +N^{2} \left(5+2(1-3N) \, {\hbox{cosech}}^{2} 
                  {\overline{\beta}}-(19+10N) \, \coth^{2} {\overline{\beta}} 
                  \right)  {\overline{\beta}}^{2} 
                  {\hbox{cosech}}^{2} {\overline{\beta}} \, \ln {\mathcal{Z}}
                  ({\overline{\beta}}) \nn \\
             & &  +2N^{2} \left((3+7N) \coth^{2} {\overline{\beta}} -5N 
                  - 1+(3-N) {\hbox{cosech}}^{2} {\overline{\beta}}\right) 
                   {\overline{\beta}}^{3} \coth {\overline{\beta}} \,
                  {\hbox{cosech}}^{2} {\overline{\beta}}  \ln {\mathcal{Z}}
                  ({\overline{\beta}}) \nn \\
             & &  + \frac{N^{3}}{2 {\overline{\beta}}^{2}} \left(\ln 
                  {\mathcal{Z}}({\overline{\beta}})\right)^{2} - \frac{N^{3}}
                  {2}  {\hbox{cosech}}^{2} {\overline{\beta}} \left(\ln 
                  {\mathcal{Z}}({\overline{\beta}})\right)^{2} + 3 N^{3} \,
                  {\overline{\beta}} \coth {\overline{\beta}} \,
                  {\hbox{cosech}}^{2} {\overline{\beta}} \,  \left(\ln 
                  {\mathcal{Z}}({\overline{\beta}})\right)^{2}  \nn \\
             & &  + N^{3} \left(1-3 \coth^{2} {\overline{\beta}} \right)
                  {\hbox{cosech}}^{2} {\overline{\beta}}
                  \left(\ln {\mathcal{Z}}({\overline{\beta}})\right)^{2}. 
\label{ms_3c_ct}
\eea
The specific heat in the third constraint framework is obtained via the 
corresponding internal energy:
\beq
C_{q}^{(3)} (\beta)      \equiv \frac{\partial U_{q}^{(3)}} {\partial T}
                         = k \; \frac{{\hat{\beta}}^{2}}{\mathfrak{N}} \left(
                          \left( \frac{N}{\overline{\beta}} \right)^{2}
                          -N {\hbox{cosech}}^{2} {\overline{\beta}} + 
                          (1-q) \;  \vartheta_{1} + (1-q)^{2} \;
                          \vartheta_{2} + \ldots\right).
\label{sh_3c_ps}
\eeq

\par

The thermodynamic quantities in the weak field limit ${\hat{\beta}} \ll 1$ 
follows directly: 
\beq
U_{q}^{(3)} (\beta) = - \frac{N_{\rm{eff}}}{3} \beta \mu^{2} H^{2}, \qquad
C_{q}^{(3)} = \frac{N_{\rm{eff}}}{3} \, k \; \hat{\beta}^{2},
\label{Ie_3c_wf}
\eeq
where the $q$-dependent effective number of spins $N_{\rm{eff}}$ reads
\beq
N_{\rm{eff}} = \frac{N}{\mathfrak{N}} 
               \left(1-(1-q)(1+N \ln2)- \frac{(1-q)^{2}}{2} 
               \left(N - 2 N \ln2 + N^{2} - (N \ln2)^{2}\right) +\ldots\right).
\label{Neff_def}
\eeq
Substituting the value (\ref{N_scale}) of the $q$-dependent scale factor 
$\mathfrak{N}$ in (\ref{Neff_def}), we notice
\begin{figure}
\begin{center}
\resizebox{120mm}{!}{\includegraphics{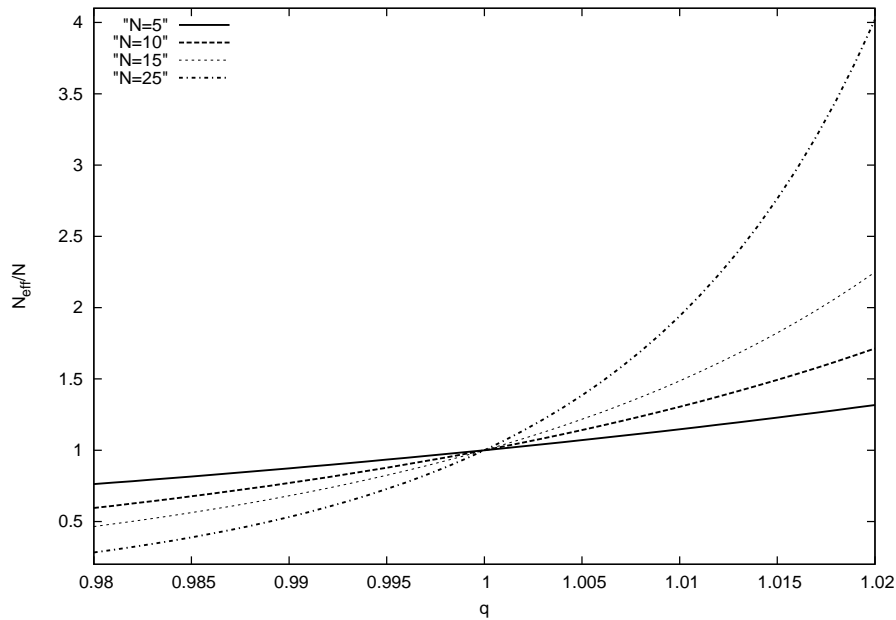}}
\caption{Dependence of the ratio $N_{\rm{eff}}/N$ on $q$ for various values of 
         $N$.}
\label{Neffective}
\end{center}
\end{figure}
that up to the perturbative order $(1-q)^{2}$ it follows 
$N_{\rm{eff}} > N$ ($N_{\rm{eff}} < N$) for the region $ q > 1$ ($q < 1$).
This is evident from the Fig. {\ref{Neffective}}.

\par
  
Turning to the magnetization (\ref{mg_3c_ps}) in a weak field  
${\hat{\beta}} \ll 1$ regime, we obtain
\beq
M_{q}^{(3)} = \frac{N_{\rm{eff}} \mu^{2} H}{3 k T},
\label{mg_3c_wf}
\eeq
where the corresponding susceptibility may be viewed as a nonextensive 
generalization of Curie's law:
\beq
\chi_{q}^{(3)}   = \frac{{\mathsf{C}}_{\rm {eff}}}{T}, 
\qquad                  
 {\mathsf{C}}_{\rm{eff}} =   \frac{N_{\rm{eff}} \mu^{2}}{3k }.
\label{ms_3c_wf}
\eeq
The thermodynamic quantities evaluated above (\ref{Ie_3c_wf}-\ref{ms_3c_wf}) 
show a nonlinear dependence on the number of spins $N$, and 
the nonextensivity parameter $(1 - q)$. This effect of nonextensivity manifest
in the aforesaid inequality between $N_{\rm{eff}}$ and $N$ 
leads to the phenomenon of dark magnetism discussed in [\cite{PPT}].

\par

The specific heat, the internal energy and the magnetization have also been 
evaluated using the approach discussed in [\cite{TMP}], where the sum of the 
$q$-weights plays a central role.  The relation between the ensemble 
probabilities (\ref{ep_eq_rel}) enables us to obtain 
${\mathfrak{c}}^{(3)}(\beta)$ as a perturbative series
\beq
{\mathfrak{c}}^{(3)}(\beta) = {\mathfrak{N}} \; \left(1 + (1-q) \;
                             {\mathfrak{P}}_{1}
                             + (1-q)^{2} \; {\mathfrak{P}}_{2} + (1-q)^{3}
                             \; {\mathfrak{P}}_{3}
                             + \ldots\right),
\label{ep_3c_wf}
\eeq
where the coefficients may be listed as
\bea
{\mathfrak{P}}_{1} &=&  N \; \ln {\mathcal{Z}}(\overline{\beta}) 
                        - N \; ({\overline{\beta}} \; \coth {\overline{\beta}}
                        -1) , \nn \\
{\mathfrak{P}}_{2} &=& \frac{N^{2}}{2} \left(\ln {\mathcal{Z}}
                       (\overline{\beta}) + 4 N^{2} - 2\,{\overline{\beta}}^{2}
                       {\hbox{cosech}}^{2} {\overline{\beta}} - 2 \,
                       {\overline{\beta}} \coth {\overline{\beta}}\right)
                       \ln {\mathcal{Z}} (\overline{\beta}) \nn \\
                   & & +\frac{N^{2}}{2} \left(4-4 {\overline{\beta}} 
                       \coth {\overline{\beta}} - {\overline{\beta}}^{2} 
                       \right) + N(1+N) {\overline{\beta}}^{3} \, \coth 
                       {\overline{\beta}} \, {\hbox{cosech}}^{2}
                       {\overline{\beta}} \nn \\
                   & & \frac{N}{2}(1+N) {\overline{\beta}}^{2} 
                       {\hbox{cosech}}^{2} {\overline{\beta}}, \nn \\
{\mathfrak{P}}_{3} &=& 5 N^{3} - \frac{N^{3}}{2} {\overline{\beta}} \;
                       ({\overline{\beta}}+8 \coth {\overline{\beta}})
                       +\frac{N^{2}}{2} (1-2N-6N{\overline{\beta}}^{2}) \;
                       {\overline{\beta}}^{2} {\hbox{cosech}}^{2} 
                       {\overline{\beta}} \nn \\
                   & & + \frac{N}{12}(17-12N+24 N^{2})
                       {\overline{\beta}}^{4} {\hbox{cosech}}^{4}
                       {\overline{\beta}} - 2N (1+N+N^{2}) 
                       {\overline{\beta}}^{5} \coth {\overline{\beta}} \;
                       {\hbox{cosech}}^{2} {\overline{\beta}} \nn \\
                   & & -N(3-N^{2}) {\overline{\beta}}^{5} \coth^{3}
                       {\overline{\beta}}  \;{\hbox{cosech}}^{2} 
                       {\overline{\beta}} - N^{2}(1+3N) {\overline{\beta}}^{5}
                       \coth {\overline{\beta}} \; {\hbox{cosech}}^{4} 
                       {\overline{\beta}} \nn \\
                   & & -\frac{N^{3}}{6} {\overline{\beta}}^{3} 
                       \coth^{3} {\overline{\beta}} - \frac{N}{4}
                       (17-23N-12N^{2}) {\overline{\beta}}^{4} 
                       \coth^{2} {\overline{\beta}} \; {\hbox{cosech}}^{2}
                       {\overline{\beta}} + 4N^{3} \ln {\mathcal{Z}}
                       ({\overline{\beta}}) \nn \\
                   & & -N^{2}\left(1-2N\right) 
                       {\overline{\beta}}^{2} {\hbox{cosech}}^{2}
                       {\overline{\beta}} \; \ln {\mathcal{Z}}
                       ({\overline{\beta}})-3N^{3} \; {\overline{\beta}} \; 
                       \coth {\overline{\beta}} \,
                       \ln {\mathcal{Z}} ({\overline{\beta}}) \nn \\
                   & & -N^{2} (3+2N) {\overline{\beta}}^{3} \coth 
                       {\overline{\beta}} \, {\hbox{cosech}}^{2} 
                       {\overline{\beta}} \, \ln {\mathcal{Z}} 
                       ({\overline{\beta}}) + \frac{N^{3}}{2} 
                       {\overline{\beta}}^{2} \, \coth^{2} {\overline{\beta}}
                       \ln {\mathcal{Z}} ({\overline{\beta}}) \nn \\
                   & & - N^{2}(3-N) {\overline{\beta}}^{4}
                       \coth^{2} {\overline{\beta}} \, {\hbox{cosech}}^{2}
                       {\overline{\beta}} \, \ln {\mathcal{Z}}
                       ({\overline{\beta}}). 
\label{ep_coeff}
\eea
The exponent property (\ref{ep_gpf}) in conjunction with our perturbative 
evaluation (\ref{ep_3c_wf}) of the sum of the $q$-weights, now allows us to 
employ the formulation specified in [\cite{TMP}] as a consistency check on 
our results. Examining  the generalized partition function a differential 
equation involving the internal energy has been established in [\cite{TMP}]:  
\beq
\beta \frac{\partial U_{q}^{(3)}}{\partial \beta} = 
\frac{\partial}{\partial \beta}  \ln_{q} {\bar{Z}}_{q}^{(3)}(\beta)
= \frac{\partial}{\partial \beta} \frac{{\mathfrak{c}}^{(3)}(\beta)-1}
 {1-q},
\label{ie_sh}
\eeq 
where  we have used the exponent relation (\ref{ep_gpf}) in the last equality. 
Substituting the internal energy that may be readily obtained via the equations
(\ref{ie_mg_rel}, \ref{mg_3c_ps}), and the sum of the 
$q$-weights given in  (\ref{ep_3c_wf}), 
it may be explicitly verified that the differential equation (\ref{ie_sh})
holds order by order in our perturbation theory. This is a nontrivial 
consistency check on our results for physical quantities. It has been noted 
earlier [\cite{CCN2}] that, as a consequence of the exponent relation 
(\ref{ep_gpf}), evaluations of the thermodynamic quantities such as specific 
heat, say, up to the second order in $(1-q)$ necessitate computing the sum of 
the $q$-weights ${\mathfrak{c}}^{(3)}(\beta)$ till the  \textit {third} order. 
That counting has been maintained in (\ref{ep_3c_wf}).

\par

Furthermore, the differential equation (\ref{ie_sh}) may be used for a direct 
extraction of the susceptibility in our model. As suggested by (\ref{mg_3c_ps})
we postulate the general functional dependence of magnetization as 
\beq
M_{q}^{(3)} = \mu \; f({\overline{\beta}}).
\label{mg_gf}
\eeq
The relations (\ref{ie_mg_rel}), (\ref{ie_sh}) and (\ref{mg_gf}) now readily 
produce the magnetic susceptibility:
\beq
\chi_{q}^{(3)} = - \frac{\mu}{{\mathfrak{N}} \; H} \; \frac{\partial}
                 {\partial {\overline{\beta}}}
                 \left(\frac{{\mathfrak{c}}^{(3)}-1}{1-q}\right).
\label{ms_gf}
\eeq 
The derivation following from (\ref{ms_gf}) precisely agrees with the magnetic 
susceptibility obtained earlier in (\ref{ms_3c_ps}). This confirms the validity
of our perturbative procedure.

%
%
%
%
%
%%%%%%%%%%%%%%%%%%%%%%%%%%%%%%%%%%%%%%%%%%%%%%%%%%%%%%%%%%%%%%%%%%%%%%%%
%
% Mean field model with temperature dependent effective field coeffecient
%
%%%%%%%%%%%%%%%%%%%%%%%%%%%%%%%%%%%%%%%%%%%%%%%%%%%%%%%%%%%%%%%%%%%%%%%%
%
\setcounter{equation}{0}
\section{Mean field model: temperature dependent \\
         effective field coefficient} 
\label{MFM}
Physical reasoning tells us that effects of nonextensivity is likely to be 
pronounced for systems embodying long-range interactions between the 
constituents. For an interacting spin system a good first approximation is 
provided by the mean field model where the long-range component of the 
interaction between the spins is taken into account via a homogeneous 
magnetic field $(H_{m})$ that is assumed to be directly proportional to the 
magnetization per spin. The Hamiltonian of the system reads
\beq
E =  - \mu (H+H_{m})  \sum_{i = 1}^{N} \cos \theta_{i},
\label{mf_eev}
\eeq
where $H_{m}$ is the dynamical field resulting from the long range 
interactions between the spins as envisaged in the mean field model. 
The generalized partition function in the third constraint reads 
\beq
{\bar{Z}}_{q}^{(3)}\left( \beta\right) = \exp_{q}\left( \varepsilon 
                                 \right)
                                 \int_{\theta_{i}=0}^{\pi}
                                 \int_{\phi_{i}=0}^{2\pi}
				 \left[ 1+(1-q) {\widetilde{\beta}} 
                                 \sum_{i=1}^{N} 
                                 \cos \theta_{i} \right]^{\frac{1}{1-q}}
                                 \prod_{i=1}^{N}
				 \sin \theta_{i}\, \hbox{d} \theta_{i}  \
				 \hbox{d} \phi_{i},
\label{gpf_mf_3c}
\eeq
where $\varepsilon = \frac{\beta  U_{q}^{(3)}}{{\mathfrak{c}}^{(3)}}$,
and the scaled dimensionless variable ${\widetilde{\beta}}$ is given by
\beq
{\widetilde{\beta}} = \frac{\beta \mu (H+H_{m})}
                      {{\mathfrak{c}}^{(3)}(\beta) + (1-q)\, \beta \,
                      U_{q}^{(3)}}.
\label{bt_3c}
\eeq
In the regime $|1-q| < 1$ we assume that the integrand is real and positive 
in the entire phase space, and obtain the generalized partition function as 
follows: 
\beq
{\bar{Z}}_{q}^{(3)}\left( \beta\right) = (2\pi)^{N} \;\;\Phi\;\;
                                         \exp_{q}\left(\varepsilon\right)
                                         {\widetilde{\beta}}^{-N} \; \;
				 {\mathfrak{S}}_{1}({\widetilde{\beta}}). 
\label{gpf_fe_3c}
\eeq

\par

The primary definition of a thermodynamic observable (\ref{q_O_3}) leads 
to the following integral form of the internal energy in the third constraint
picture: 
\bea
U_{q}^{(3)}(\beta) &=& - \, \mu (H+H_{m})
                       \left( {{\bar{Z}}_{q}^{(3)}(\beta)} \right)^{-1}  \;  
                       \left(\exp_{q}\left(\varepsilon\right)\right)^{q}
                       \nn \\
                   & & \int_{\theta_{i}=0}^{\pi}
                       \int_{\phi_{i}=0}^{2\pi}
                       \sum_{i=1}^{N} \cos \theta_{i} \;
 	               \left[ 1+(1-q) {\widetilde{\beta}} 
                       \sum_{i=1}^{N} \cos \theta_{i} \right]^{\frac{q}{1-q}}
                       \prod_{i=1}^{N} \sin \theta_{i}\, \hbox{d} \theta_{i}
		       \hbox{d} \phi_{i}.
\label{Ie_mf_3c}
\eea
Implementing the above phase space integration we obtain the internal energy as
\beq
U_{q}^{(3)} (\beta) = \left(\frac{2 \pi}{{\widetilde{\beta}}}\right)^{N} \;
                      \Phi \; \mu (H+H_{m}) \;
                      \left( N  {\widetilde{\beta}}^{-1}\; 
		      {\mathfrak{S}}_{1}(\widetilde{\beta})
		      - {\mathfrak{S}}^{\prime}_{1}(\widetilde{\beta})\right)\;
                      \frac{\left(\exp_{q}\left(\varepsilon\right)\right)^{q}}
                      {{\bar{Z}}_{q}^{(3)}}.
\label{Ie_3c_aI}
\eeq
On substituting the generalized partition function (\ref{gpf_fe_3c}) the 
implicit equation (\ref{Ie_3c_aI}) may be recast as follows:
\beq
U_{q}^{(3)} (\beta) = \frac{\mu (H+H_{m})}{1+(1-q) \, \varepsilon} 
                      \left( \frac{N}{{\widetilde{\beta}}} -
                      \frac{ {\mathfrak{S}}^{\prime}_{1}(\widetilde{\beta})}
                      {{\mathfrak{S}}_{1}(\widetilde{\beta})} \right).
\label{Ie_3c_fe}
\eeq
We expand the term in the parenthesis up to second order in the nonextensivity
parameter $(1-q)$ and fourth order in the dynamical variable 
$\beta \mu (H+H_{m})$.
The nonzero contributions in the resulting expansion read
\beq
U_{q}^{(3)}(\beta) = - \frac{N\, \beta \mu^{2} (H+H_{m})^{2}}
                       {3 \; {\mathfrak{c}}^{(3)}(\beta)}
                       \left({\mathcal{U}}_{1} - 
                       \frac{{\mathcal{U}}_{2}}{15} \;\;
                       \left(\frac{\beta \mu (H+H_{m})}
                       {{\mathfrak{c}}^{(3)}(\beta)}\right)^{2} 
                       + \ldots \right),
\label{Ie_ps_3c}
\eeq
where the perturbative coefficients are given by
\bea
{\mathcal{U}}_{1} &=&  1 - (1-q)\,
                       \left(1 + 2 \varepsilon\right) 
                      + (1-q)^{2} \, \varepsilon
                       \left(2 + 3 \, \varepsilon\right)+ \ldots,  \nn \\
{\mathcal{U}}_{2} &=& 1 - (1-q) \, \left(6-10N+4 \, \varepsilon\right) \nn \\
                  & & +(1-q)^{2} 
                       \left(11-25N+24 \varepsilon 
                       - 40 \varepsilon + 10N \varepsilon^{2}\right) + \ldots.
\label{Ie_pt_3c}
\eea
The phase space integral corresponding to the sum of the $q$-weights may be 
read off via (\ref{ep_3c_def}) and (\ref{ep_gpf}):
\beq
{\mathfrak{c}}^{(3)} (\beta) = \left(\frac{
                               \exp_{q}\left(\varepsilon \right)}
                               {{\bar{Z}}^{(3)}_{q}(\beta)}\right)^{q} 
                               \int_{\theta_{i}=0}^{\pi}
                               \int_{\phi_{i}=0}^{2\pi}
	                       \left[ 1+(1-q) {\widetilde{\beta}} 
                               \sum_{i=1}^{N} \cos \theta_{i} \right]
                               ^{\frac{q}{1-q}}
                               \prod_{i=1}^{N} \sin \theta_{i}\, 
                               \hbox{d} \theta_{i}  \, \hbox{d} \phi_{i}.
\label{qw_mf_3c}
\eeq
Performing the above integrations and subsequently substituting the 
generalized partition function (\ref{gpf_fe_3c}) above sum of the $q$-weights
assumes the form 
\beq
{\mathfrak{c}}^{(3)}(\beta) = {\mathfrak{N}} \;\; 
                              \Lambda \;\;
                              {\widetilde{\beta}}^{(q-1)N} \;\; 
                              {\mathfrak{S}}_{2}({\widetilde{\beta}}) \;\;
                              ({\mathfrak{S}}_{1}({\widetilde{\beta}}))^{-q}.
\label{qw_3c_fe}
\eeq
As done before in the instance of the internal energy in (\ref{Ie_ps_3c})
the rhs of the expression (\ref{qw_3c_fe}) is expanded perturbatively up to 
second order in the nonextensivity parameter $(1-q)$ and  fourth order in the 
variable $\beta \mu (H+H_{m})$:
\bea
{\mathfrak{c}}^{(3)}(\beta) &=& {\mathfrak{N}} \bigg(1+ (1-q) N \ln2 +
                                (1-q)^{2} \, \frac{N}{2} \left(1+N+N(\ln2)^{2}
                                \right) \nn \\
                            & & + \wp_{1}(\varepsilon) \left(\frac{\beta \mu 
                                (H+H_{m})} 
                                {{\mathfrak{c}}^{(3)}(\beta)}\right)^{2}
                                + \wp_{2}(\varepsilon) \left(\frac{\beta \mu 
                                (H+H_{m})} 
                                {{\mathfrak{c}}^{(3)}(\beta)}\right)^{4} 
                                + \ldots\bigg),
\label{qw_3c_ps}
\eea
where the perturbative expansion of the coefficients of monomials of the
field variable read
\bea
\wp_{1}(\varepsilon) &=& - \frac{N}{6}\left((1-q) + (1-q)^{2}  \left(1-N \ln2
                         +2 \; \varepsilon \right)+ \ldots\right), \nn \\
\wp_{2}(\varepsilon) &=& \frac{N}{60} \left((1-q)-(1-q)^{2} 
                         \left(6-\frac{15N}{2}-N \ln2
                         +4 \; \varepsilon \right)+ \ldots\right).
\label{qw_pt}
\eea

\par

\begin{figure}
\begin{center}
\resizebox{120mm}{!}{\includegraphics{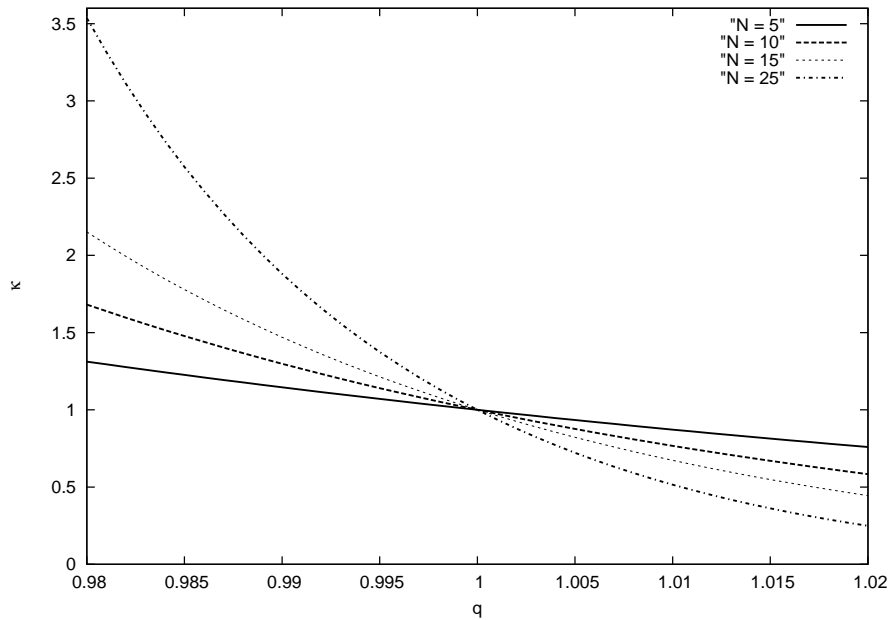}}
\caption{Variation of the quantity $\kappa$ with $q$ for various values of 
         $N$.}
\label{critical_temperature}
\end{center}
\end{figure}

Towards obtaining the magnetization we first explicitly obtain the quantities 
$U_{q}^{(3)}(\beta)$ and ${\mathfrak{c}}^{(3)}(\beta)$ by solving the pair of 
simultaneous implicit equations (\ref{Ie_ps_3c}) and (\ref{qw_3c_ps}). We can 
systematically obtain their solutions in an order by order
perturbation theory where we retain terms up to $(1-q)^{2}$ in the 
nonextensivity parameter and $(\beta \mu (H+H_{m}))^{4}$ in the field 
variable. The relevant expression for the internal energy reads 
\beq
U_{q}^{(3)}(\beta) = - \frac{N}{3} \frac{\beta \mu^{2} (H+H_{m})^{2}} 
                       {\mathfrak{N}}\;\; \Gamma + \frac{N}{45} 
                       \frac{\beta^{3} \mu^{4} (H+H_{m})^{4}}
                       {{\mathfrak{N}}^{3}} \;\; \Xi + \ldots,
\label{Ie_ee}
\eeq
where the $q$-dependent numerical factors $\Gamma$ and $\Xi$ are given by 
\bea
\Gamma &=& 1-(1-q) \left(1+N \ln2\right) \nn \\
       & &   - (1-q)^{2}\; \frac{N}{2}
           \left(1 + N -2 \, \ln2 - N \, (\ln2)^{2} \right) 
           + \ldots, \\
\Xi &=& 1-(1-q) \left(6+\frac{5N}{2}+3N \, \ln2\right) \nn \\
    & & + (1-q)^{2} \; 
        \left(11- \frac{3N}{2}(N+1) +\frac{N \ln2}{2} \;
         (36+15N+9N \ln2)\right)
         + \ldots. 
\label{qterms}
\eea
The perturbative expansion also yields the sum of $q$-weights as
\bea
{\mathfrak{c}}^{(3)}(\beta) &=&{\mathfrak{N}} \bigg(1+ (1-q) N \ln2 +
                                (1-q)^{2} \, \frac{N}{2} \left(1+N+N(\ln2)^{2}
                                \right) \nn \\
                            & & + {\varPi}_{1} 
                                \left(\frac{\beta \mu 
                                (H+H_{m})} {{\mathfrak{N}}}\right)^{2}
                                +{\varPi}_{2} 
                                \left(\frac{\beta \mu 
                                (H+H_{m})} {{\mathfrak{N}}}\right)^{4} 
                                + \ldots\bigg),
\label{qw_ee}
\eea
where the coefficients may be listed as follows:
\bea
{\varPi}_{1} &=& - \frac{N}{6}\left((1-q) + (1-q)^{2}  \left(1-N \ln2
                        \right)+ \ldots\right), \nn \\
{\varPi}_{2} &=& \frac{N}{60} \left((1-q)-(1-q)^{2} 
                         \left(6+\frac{5N}{2}+3N \ln2\right)+ \ldots\right).
\label{qw_ee_pt}
\eea
As the internal energy and the magnetization are related via the standard
thermodynamic expression (\ref{ie_mg_rel}) we may now readily obtain a 
perturbative expansion for the magnetization by employing the corresponding 
series (\ref{Ie_ee}) for the internal energy:
\beq
M_{q}^{(3)} = \mu \; \left(\frac{N}{3 \; {\mathfrak{N}}} \; 
              \beta \mu (H+H_{m}) \; \Gamma
              - \frac{N}{45 \; {\mathfrak{N}}^{3}} \;
              \left(\beta \mu (H+H_{m})\right)^{3} \; \Xi\right).
\label{mg_ps_3c}
\eeq
The homogeneous magnetic field $H_{m}$ is taken to be proportional to the
magnetic moment per spin.  
To account for the quantum effect of spin fluctuations the 
proportionality factor, known as the effective field coefficient 
$(\lambda (T))$, is considered to be temperature dependent 
[\cite{SW},\cite{ZXXZ}]. Following [\cite{ZXXZ}] we here study a mean field 
spin model with the field coefficient  $\lambda(T)$ being linearly 
dependent on temperature. The above discussion leads to the field variable
\beq
H_{m} = \lambda (T) \;m_{q}^{(3)}, \qquad
m_{q}^{(3)} \equiv \frac{M_{q}^{(3)}}{N}, \qquad
\lambda (T) = \xi + \zeta \; T,
\label{mf_mg_L}
\eeq
where the coefficients $\xi$ and $\zeta$ characterize the long range spin 
interaction. Substituting (\ref{mf_mg_L}) in (\ref{mg_ps_3c}) and considering 
the vanishing limit of the the external field $H = 0$, the magnetization reads
\beq
m_{q}^{(3)} \ \Theta
\left(\frac{3 \,\Xi}{5 \,\Gamma^{3} \, \mu^{2}} 
\left(1+\frac{\zeta T}{\xi}\right)^{3} \
\Theta^{2} \ \left(\frac{T_{c}^{(3)}}{T}\right)^{3} 
\Big(m_{q}^{(3)}\Big)^{2} 
+1-\frac{T_{c}^{(3)}}{T}\right) = 0,
\label{mg_ce}
\eeq
where the critical temperature is 
\beq
T_{c}^{(3)} = \frac{\xi \, \mu^{2} \, \Gamma}{3k \, {\mathfrak{N}}\; \Theta},
\qquad
\Theta = 1 - \frac{\zeta \; \mu^{2} \; \Gamma}{3k \; {\mathfrak{N}}}.
\label{Tc_3c}
\eeq
To visualize the variation of the critical temperature with respect to the 
nonextensivity parameter $(1-q)$, we, in Fig.\ref{critical_temperature}, plot 
the ratio $\kappa$ defined as  
\beq
\kappa \equiv \frac{\beta_{c}^{(3)}+\delta}
                      {\beta_{c}^{(3)}|_{q=1}+\delta}, \qquad
\delta = \frac{\zeta}{\xi \, k}.
\label{kpa_def}
\eeq
From Fig.\ref{critical_temperature} it may be inferred that the critical 
temperature increases (decreases) compared to its standard Boltzmann-Gibbs 
value in the regime $q>1$ ($q<1$). This result is qualitatively similar to the 
observation in [\cite{RAAO}] where these authors have adopted an alternate 
definition of the temperature for the nonextensive spin system. The mean field 
model studied by these authors is slightly different from the one considered 
here in that we assume the dynamical field strengh to be temperature dependent 
in order to accommodate the quantum fluctuations in spin configurations.   
\par

Equation (\ref{mg_ce}) suggests that above the critical temperature 
$T>T_{c}^{(3)}$, the only real solution for the magnetization in null 
external field condition is given by
\bea
m_{q}^{(3)}|_{H = 0}=0
\label{m_para}
\eea
that corresponds to the paramagnetic phase. Using the standard definition 
(\ref{ms_def}) in conjunction with the relations (\ref{mg_ps_3c}) and 
(\ref{m_para}) we now obtain the
magnetic susceptibility in the paramagnetic regime:
\beq
\chi_{q}^{(3)} = \frac{{\mathcal{C}}}{T-T_{c}^{(3)}}, \qquad
{\mathcal{C}} = \frac{N \, \mu^{2} \, \Gamma}
                {3k \, {\mathfrak{N}} \; \Theta}.
\label{cw_3c}
\eeq
It is evident from (\ref{mg_ce}) that below the critical temperature 
$T < T_{c}^{(3)}$ the magnetization in the null external field limit 
is given by two stable real values given by 
\beq
m_{q}^{(3)}|_{H=0} = \pm \;  \mu \; \sqrt{\frac{5 \, 
              \Gamma^{3}}{3 \, \Xi}} \; \;  
              \Theta^{-1}
              \left(1+\frac{\zeta}{\xi}T\right)^{-3/2}
              \frac{T}{T_{c}^{(3)}} \;
              \sqrt{1-\frac{T}{T_{c}^{(3)}}}.
\label{mg_3c}
\eeq
This corresponds to the ferromagnetic transition as observed in the 
generalized mean field approach in the nonextensive framework.
In the regime $T \rightarrow T_{c}^{(3)\;-}$ the susceptibility in the 
null external
field limit may be obtained via (\ref{mg_ps_3c}) and (\ref{mg_3c}):  
\beq
\chi_{q}^{(3)} = \frac{{\mathcal{C}}/2} {T_{c}^{(3)} - T}.
\label{sus_less}
\eeq 
As characteristic of the mean field approach, the divergence of the 
susceptibility follows critical exponent law 
\beq
\chi_{q}^{(3)} \sim \frac{1}{|T-T_{c}^{(3)}|},
\label{sus_div}
\eeq
where the critical temperature depends on the nonextensivity parameter
$(1-q)$.  Of course in the extensive $q \rightarrow 1$ limit, the standard 
physical quantities corresponding to the Boltzmann-Gibbs statistics are 
recovered.  We also quote the results obtained in the usual mean field model, 
where the dynamical proportionality factor $\lambda$ is regarded as 
independent of temperature \textit{i.e.} the linear coefficient in 
(\ref{mf_mg_L}) assumes the value $\zeta = 0$. In this limit the critical 
temperature and the Curie-Weiss constant in the nonextensive scenario may be 
arrived at via  
(\ref{Tc_3c}) and (\ref{cw_3c}): 
\beq
{\mathcal{T}}_{c}^{(3)} = \frac{\xi \, \mu^{2} \, \Gamma}{3k \, 
                          {\mathfrak{N}}},
\qquad
{\mathfrak{C}} = \frac{N \, \mu^{2} \, \Gamma}
                {3k \, {\mathfrak{N}}}.
\label{Tc_cl}
\eeq
%
%
%%%%%%%%%%%%%%%%%%%%%%%%%%%%%%%%%%%%%%%%%%%%%%%%%%%%
%
%    Remarks
%
%%%%%%%%%%%%%%%%%%%%%%%%%%%%%%%%%%%%%%%%%%%%%%%%%%%%
%
\setcounter{equation}{0}
\section{Remarks} 
\label{REM}
Our main focus in the present work has been to study a system of spins 
with long range interactions approximated by a mean field model governed by 
the nonextensive Tsallis statistics. To incorporate the quantum spin 
fluctuations the mean field model investigated here is assumed 
to have the dynamical field 
strength coefficient depending linearly on temperature. The nonextensivity is 
implemented by using a perturbative technique, where the implicit simultaneous
equations involving the internal energy and the sum of $q$-weights were solved 
explicitly as series expansions up to the order $(1-q)^{2}$ in the 
nonextensivity parameter. The perturbation method developed here may be 
continued to an arbitrary order in the parameter \,$(1-q)$. The signature 
of the nonextensivity is evident 
as the critical temperature is found to depend on the number of spins $N$ and 
the deformation variable \,$(1-q)$. Compared to its standard Boltzmann-Gibbs 
value the critical temperature increases (decreases) for the domain 
$q > 1\;\; (q < 1).$ The extended Curie-Weiss law characterizing 
the susceptibility in the regions above and below the critical temperature 
has been generalized to the nonextensive case. Analogous to the critical 
temperature the Curie-Weiss constant 
also embodies the effects of nonextensivity. Another interesting feature 
reflecting nonextensivity emerges for noninteracting spins in the 
presence of a background field, where we observe the presence of 
dark magnetism. This supports the results [\cite{PPT},\cite{MPP}] obtained 
earlier. Parallel to the observation in [\cite{MPP}] our analysis indicates 
that the effective number of 
spins $N_{\rm{eff}} > N\;(N_{\rm{eff}} < N)$ for the regime $q>1$\; ($q<1$). 

\par

The mean field technique used in the current work may be fruitfully
employed in calculating the magnetic properties of systems which form
clusters [\cite{RVC}].  A cluster is typically a macroscopic region 
consisting of a large number of spins whose interactions are described 
via locally homogeneous mean fields in the domain of each cluster. Then 
each cluster is approximated by a single effective spin forming an ensemble 
of varying spins whose magnetic properties may be studied by adopting a 
suitable model. The results obtained here may facilitate studying such models 
in the context of Tsallis statistics.

%
%
%
%
%
%
%%%%%%%%%%%%%%%%%%%%%%%%%%%%%%%%%%%%%%%%%%%%%%%%%%%
%
%    Acknowledgement
%
%%%%%%%%%%%%%%%%%%%%%%%%%%%%%%%%%%%%%%%%%%%%%%%%%%%
%
\section*{Acknowledgement} 
The work of R. Chandrashekar is supported by a fellowship offered by 
Council of Scientific and Industrial Research (India).

%
%%%%%%%%%%%%%%%%%%%%%%%%%%%%%%%%%%%%%%%%%%%%%%%%%%%%
%
%    References
%
%%%%%%%%%%%%%%%%%%%%%%%%%%%%%%%%%%%%%%%%%%%%%%%%%%%%
%

%
\end{document}